\documentclass[graybox]{svmult}
\usepackage{makeidx}  
\usepackage{graphicx} 
\usepackage{multicol}        
\usepackage{braket}
\usepackage{amsmath}
\usepackage{amssymb}
\usepackage{dsfont}
\usepackage{hyperref}
\usepackage{wrapfig}

\usepackage[bottom]{footmisc}

\usepackage{newtxtext}       %
\usepackage[varvw]{newtxmath}       

\usepackage[backend=bibtex,
            style=phys,
            maxnames=3,
            biblabel=brackets, 
            defernumbers]{biblatex}
\AtBeginBibliography{\small}
\usepackage[explicit]{titlesec}
\titleformat{name=\paragraph,numberless}[runin]
  {\normalfont\normalsize\bfseries}{}{0pt}{#1.}
\titlespacing{\paragraph}{%
  0pt}{
  0.5\baselineskip}{
  1em}

\makeindex             

\title*{Simulating dynamics of correlated matter with neural quantum states}
\author{Markus Schmitt\orcidID{0000-0003-2223-8696} and Markus Heyl\orcidID{0000-0002-7126-1836}}
\institute{Markus Schmitt 
\at University of Regensburg, 93053 Regensburg, Germany, \email{markus.schmitt@ur.de}
\at Forschungszentrum J\"ulich GmbH (PGI-8), 52425 J\"ulich, Germany
\and Markus Heyl \at Theoretical Physics III, Center for Electronic Correlations and Magnetism,
Institute of Physics, University of Augsburg, D-86135 Augsburg, Germany, \email{markus.heyl@uni-a.de}}
\begin{document}

\maketitle

\abstract*{Same abstract text here for some technical reason...}

\abstract{While experimental advancements continue to expand the capabilities to control and probe non-equilibrium quantum matter at an unprecedented level, the numerical simulation of the dynamics of correlated quantum systems remains a pivotal challenge -- especially in intermediate spatial dimensions. Neural quantum states are emerging as a new computational tool to investigate the time evolution of many-body quantum systems in previously inaccessible regimes. We review the recent progress in the field with a focus on the different time propagation methods, an overview of the reported applications, and a discussion of the major current challenges.}

\section{Introduction}

Impressive experimental advancements over the past two decades have led to unprecedented control of quantum matter enabling to observe their dynamics with unique detail and precision.
Thereby, exciting avenues have been opened up for the exploration of novel genuinely non-equilibrium physics at an unparalleled level.
For quantum simulation and computing devices the immense potential to probe fundamental dynamical principles of quantum matter has already been demonstrated multiple times, including the observation of quantum many-body scars \cite{Bernien2017}, many-body localization \cite{Schreiber2015,Smith2016, Bordia2017}, dynamical quantum phase transitions \cite{Jurcevic2017}, or discrete time crystals \cite{Choi2017,Zhang2017}.
In the context of condensed matter systems, modern ultrafast pump-probe techniques have enabled the experimental investigation of correlated materials beyond equilibrium at a new level.
Thereby, ultrafast manipulation of material properties becomes feasible \cite{Buzzi2018, Mazzone2021,delaTorre2021} and intriguing non-equilibrium phenomena could be uncovered, such as light-induced superconductivity \cite{Mitrano2016} or metastable phases \cite{Stojchevska2014}.

However, understanding and theoretically describing such dynamics, particularly in strongly interacting systems, represents one of the central challenges in quantum theory.
While numerical simulations are of key importance to advance at this frontier, traditional methods are often facing fundamental limitations.
Exact diagonalization (ED) suffers from the curse of dimensionality caused by the exponential growth of the Hilbert space with system size~\cite{Lin1993}.
Quantum Monte Carlo (QMC) techniques, successful in equilibrium descriptions, encounter the sign problem~\cite{Troyer2005}, and tensor network (TN) methods struggle with entanglement growth and tensor contraction complexity~\cite{Schollwoeck2011}. 
Nevertheless, within the last decades tremendous progress has been achieved specifically in the extreme cases of one spatial dimension (1D) through TNs \cite{Paeckel2019} and infinite dimensionality via dynamical mean-field theory (DMFT)~\cite{Aoki2014}.
This, however, leaves a critical gap in our ability to simulate quantum systems in intermediate dimensions two and three.
In recent years the neural quantum state (NQS) has emerged as a novel promising numerical simulation technique to tackle the quantum many-body problem \cite{Carleo2017}.
This approach leverages the remarkable capabilities of machine learning methods for quantum physics.
The key idea behind NQS is to represent the wave function, the fundamental object describing quantum physical systems, as an artificial neural network (ANN).
The NQS technique has seen impressive advances, particularly concerning the simulation of two-dimensional (2D) quantum matter, making NQS a competitive numerical method as compared to other existing numerical techniques.
Examples of these advances in simulating the dynamics in such systems are the first theoretical verification of the quantum Kibble-Zurek mechanism in 2D \cite{Schmitt2022}, emulated quantum computation with large qubit registers \cite{Medvidovic2021}, the numerically exact calculation of dynamical susceptibilities \cite{MendesSantos2023}, or the ab-initio simulation of correlated electrons out of equilibrium \cite{Nys2024}. 
These impressive developments highlight the potential of NQS to become a key numerical reference method for 2D quantum many-body systems, possibly enabling new discoveries and insights into fundamental quantum physics. 
In the following presentation, we will first outline the methodical basics of time-dependent NQS simulations in Section \ref{sec:TDVP}. In Section \ref{sec:applications} we summarize the state of the art by giving an overview over applications, that have been addressed in the existing literature. After that, in Section \ref{sec:challenges}, we will discuss the open questions and challenges that need to be addressed for the further advancement of the NQS approach.

\section{Variational time evolution with neural quantum states}
\label{sec:TDVP}

On a general level, the main idea of NQS-based time-evolution algorithms is to find an efficient representation of the wave function $\ket{\psi(t)}$ or density matrix $\rho(t)$, whose dynamics is described by a Schrödinger, von Neumann, or Lindblad equation, respectively.
For this purpose, a variational ansatz is selected for the quantum state, utilizing an ANN, where the weights and biases of the ANN serve as the set of variational parameters.
Importantly, the performance of such a variational technique relies on two pillars.
The first pillar is the expressive power of the ansatz. It needs to be capable of encoding the desired classes of wave functions with sufficient efficiency.
As the utilized ANNs are universal function approximators \cite{Cybenko1989, Hornik1991}, a sufficiently large NQS can, in principle, represent any quantum wave function. 
This renders the NQS technique a numerically exact procedure with the ANN size as the control parameter.
The second pillar is the effective optimization or training.
Even if a representation of the desired state exists within the chosen ansatz, it is a-priori not clear, whether it can be found efficiently.
Generally, training ANNs amounts to optimization in high-dimensional non-convex landscapes, which is a challenging problem.
However, NQS methods can draw from a rich toolbox for these purposes developed in both the deep learning and physics communities.
We will start by briefly recapitulating how artificial neural networks can represent (time-evolved) quantum states in Section \ref{subsec:nqs}. In Section \ref{subsec:tdvp} we will discuss different approaches to optimize NQS, in order to variationally propagate the quantum state in time.
Finally, we will summarize alternative evolution techniques in Section \ref{subsec:other_evolution}.

\subsection{Neural quantum states}
\label{subsec:nqs}
Consider a many-body Hilbert space $\mathcal H=\bigotimes_{l=1}^N\mathcal H^{(l)}$ of $N$ degrees of freedom on a lattice and a full computational basis of choice, $\{\ket{\vec x}=\ket{x_1}\otimes\ldots\otimes\ket{x_N}\}$, labelled by $\vec x=(x_1,\ldots,x_N)$.
Then, any quantum state $\ket{\psi}$ can be represented in this basis as
\begin{align}
    \ket{\psi}=\sum_{\vec x}\psi({\vec x})\ket{\vec x}\ .
\end{align}
For instance, for systems composed of $N$ spin-$1/2$ degrees of freedom a natural choice for such a basis would be the basis of spin configurations $\vec s = (s_1,\dots,s_N)$ with $s_i = \uparrow,\downarrow$.
In what follows, however, we keep the computational basis $\vec x$ of general type with the only constraint that the local Hilbert-space dimensions $\mathrm{dim}[\mathcal{H}^{(l)}]<\infty$ are all finite for simplicity but without loss of generality. See Refs.~\cite{Medvidovic2023, Nys2024} for the generalization to continuous variables.
Given a computational basis, the full information of the state is then contained in the complete set of wave function coefficients $\psi({\vec x})\in\mathbb C$.
Since the dimension of the $N$-particle Hilbert space grows exponentially with $N$, it is crucial for scalable numerical approaches to find compressed representations of the wave function. The idea of variational wave functions is to introduce efficiently tractable functions $\psi_{\vec\theta}(\vec x)$ determined by a set of parameters $\vec\theta\in\mathbb R^P$. If, for a wave function of interest with given amplitudes $\psi({\vec x})$, the parameters $\vec\theta$ can be adjusted, such that $\psi_{\vec\theta}(\vec x)\approx\psi({\vec x})\,\,\forall\vec x$, the variational ansatz together with the optimal set of parameters constitutes a compressed representation of the wave function. While it is possible to use complex-valued parameters, we will in the following consider the more general case of real parametrizations unless stated otherwise. Complex parametrizations can always be viewed as real parametrizations by considering the real and imaginary parts as independent parameters.

\begin{figure}[t]
\center
\includegraphics{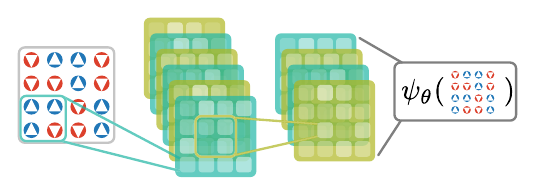}
\caption{Schematic depiction of a neural quantum state. The computational basis configuration $\vec x$ (left) is processed by the artificial neural network (center) to produce the corresponding wave function coefficient $\psi_{\vec\theta}(\vec x)\in\mathbb C$ (right). Image adapted from Ref.~\cite{MendesSantos2023}}
\label{fig:nqs_schematic}
\end{figure}
In the context of NQS we view $\psi_{\vec\theta}(\vec x)$ as a function
\begin{equation}
    \vec x \mapsto \psi_{\vec \theta}(\vec x) \in \mathbb{C}
\end{equation}
mapping configurations $\vec x$ to complex numbers, which we parametrize through an ANN.
In essence, the ANN accepts as an input a configuration $\vec x$, corresponding to the computational basis configuration, and outputs the corresponding wave function amplitude $\psi_{\vec\theta}(\vec x)$, see Fig.~\ref{fig:nqs_schematic} for an illustration.
In general, an NQS $\ket{\psi_{\vec \theta}}$ is then defined as a quantum state whose amplitudes $\psi_{\vec \theta}$ are given by an ANN,
\begin{equation}
    \ket{\psi_{\vec \theta}} = \sum_{\vec x} \psi_{\vec \theta} (\vec x) \ket{\vec x}.
\end{equation}
Within the framework presented in the following, the choice of the ANN architecture remains free in general and numerous possibilities have been explored in the literature \cite{Lange2024}. While the strengths and weaknesses of different architectures are subject of ongoing research, the incorporation of symmetries of the physical problem into the ansatz has turned out to be an important design principle, that typically enhances the performance. 

It is common practice to choose as the ANN output the logarithmic wave function coefficients $\chi_{\vec\theta}(\vec x)$, such that $\psi_{\vec\theta}(\vec x)=\exp\big(\chi_{\vec\theta}(\vec x)\big)$.
The physical reason underlying this choice is that amplitudes of many-body states are in general exponential functions of system size.
One direct way to see this property is to consider the constraint emerging from the normalization of the wave function $\sum_{\vec x} |\psi_{\theta}(\vec x)|^2=1$, where for typical wave functions exponentially many (size of the Hilbert space) nonzero numbers have to sum up to one.
This is only possible if each of the summands itself is exponentially small.
Thus, if one were to work directly with the amplitudes, problems with numerical precision can occur for large systems.
However, it turns out that representing wave function amplitudes directly is still feasible for up to $400$ spin-1/2 degrees of freedom~\cite{AoChen2024}.
If one uses the logarithmic representation via $\chi_{\vec \theta}(\vec x)$, the output of the ANN becomes linear in system size, which allows to resolve amplitudes ranging over orders of magnitude with the ANN ansatz, remaining well defined even in the thermodynamic limit.
At the same time, however, the encoding of the logarithmic wave function can lead to other numerical problems: Vanishing wave function amplitudes require a divergent ANN output, as will be discussed in more detail in Sec.~\ref{subsubsec:open_questions_infintesimal}, and the efficiency of incorporating symmetries in the ansatz may suffer \cite{Reh2023}.
Finally, the preferred strategy has to be chosen on a case-by-case basis.

\subsubsection{Monte Carlo estimation of expectation values}
\label{subsubsec:mc_estimation}

For a given NQS $\ket{\psi_\theta}$ one can efficiently calculate the expectation value of observables by means of a Monte-Carlo sampling scheme \cite{vanDenNest2011}.
Specifically, considering an operator $\hat{O}$, the corresponding expectation value can be expressed as
\begin{equation}
    \frac{\bra{\psi_{\vec \theta}} \hat{O} \ket{\psi_{\vec \theta}}}{\braket{\psi_{\vec\theta}|\psi_{\vec\theta}}} = \sum_{\vec x} \frac{\psi^\ast_{\vec \theta}(\vec x)}{\braket{\psi_{\vec\theta}|\psi_{\vec\theta}}} \bra{\vec x} \hat{O} \ket{\psi_{\vec \theta} } = \sum_{\vec x} \frac{\left| \psi_{\vec \theta}(\vec x) \right|^2}{\braket{\psi_{\vec\theta}|\psi_{\vec\theta}}} O^{\text{loc}}_{\vec \theta}(\vec x)
\end{equation}
with the so-called local estimator of $\hat O$, 
\begin{align}
    O^{\text{loc}}_{\vec \theta}(\vec x) = \frac{\bra{\vec x} \hat{O} \ket{\psi_{\vec \theta} }}{\braket{\vec x|\psi_{\vec \theta}}}=\sum_{\vec x'}\braket{\vec x|\hat O|\vec x'}\frac{\psi_{\vec\theta}(\vec x')}{\psi_{\vec\theta}(\vec x)}\ .
    \label{eq:local_estimator}
\end{align}
In this representation, the quantum expectation value is expressed as the mean of the local estimator with respect to the Born probability $p_{\vec \theta}(\vec x)\propto\left| \psi_{\vec \theta}(\vec x) \right|^2$.
Thereby, the sum over the computational basis can be efficiently estimated by means of a Monte Carlo sampling. Upon obtaining a sample $\mathcal S=\{\vec x^{(i)}\}_{i=1\ldots N_{MC}}$ from the distribution $p_{\vec \theta}(\vec x)$, the estimate (under the assumption of uncorrelated samples) is
\begin{equation}
    \frac{\bra{\psi_{\vec \theta}} \hat{O} \ket{\psi_{\vec \theta}}}{\braket{\psi_{\vec\theta}|\psi_{\vec\theta}}} = \frac{1}{N_{MC}} \sum_{\vec x\in\mathcal S}  O^{\text{loc}}_{\vec \theta}(\vec x) + \mathcal E_{\hat O}^{\mathcal S} \, .
\end{equation}
with a statistical error contribution $\mathcal E_{\hat O}^{\mathcal S}$.
According to the central limit theorem, the expected deviation of the sample mean from the exact expectation value is
\begin{align}
    \mathbb E\big[|\mathcal E_{\hat O}^{\mathcal S}|\big]=\sqrt{\frac{\text{Var}_{|\psi(\vec x)|^2}\big[O^{\text{loc}}_{\vec \theta}\big]}{N_{MC}}}\ .
    \label{eq:MC_error}
\end{align}
Hence, efficient sampling is possible, if the variance of the local estimator is bounded. 
Notice, that while the local estimator of Eq.~\eqref{eq:local_estimator} is the default choice in practice, one may need to resort to alternatives with guaranteed bounds on the variance \cite{vanDenNest2011}.

In many cases, standard Markov Chain Monte Carlo techniques are sufficient to generate samples from the Born distribution. An alternative is to choose an NQS architecture with an autoregressive property, which allows to generate uncorrelated samples by subsequent individual network evaluations~\cite{Sharir2020}.

\subsubsection{Short-time evolution}

While we will discuss the question to which extent NQS can generally represent time-evolved quantum many-body states in more detail in Sec.~\ref{subsec:challenges_expressivity}, the short-time evolution is guaranteed to be accessible by perturbative constructions.
Consider an initial condition $\ket{\psi_0} = \ket{\psi_{{\vec \theta}_0}}$ for some dynamical problem in NQS form with variational parameters $\vec \theta_0$.
This could, for instance, be the ground state of some initial Hamiltonian in a quantum quench setting or a certain product state implementable experimentally in a quantum simulator device.
Without loss of generality, let us now imagine a general time-independent Hamiltonian $H$ starting to act from time $t=0$.
Then, we can express the time-evolved state $\ket{\psi(t)}$ as
\begin{equation}
    \ket{\psi(t)} = \sum_{\vec x} \psi_{\vec \theta_0}(\vec x) \eta_{\vec \theta,t} (\vec x) \ket{\vec x}, \, \eta_{\vec \theta,t}(\vec x) = \frac{\bra{\vec x} e^{-i  \hat H t} \ket{\psi_{\vec \theta_0}}}{\braket{x|\psi_{\vec \theta_0}}} \, .
\end{equation}
For short times, we can then use a cumulant expansion to obtain~\cite{Kubo1962} $ \eta_{\theta,t}(\vec x)= e^{\mathcal{H}(\vec x, t)}$
with 
\begin{equation}
     \mathcal{H}(\vec x, t) = -i t E - \frac{t^2}{2} \Delta E^2 + \mathcal{O}(t^3)\ .
\end{equation}
Here, $E=\frac{\bra{\psi_{\vec \theta}} \hat H \ket{\psi_{\vec \theta}}}{\braket{\psi_{\vec \theta}|\psi_{\vec \theta}}}$ and $\Delta E^2 = \frac{\bra{\psi_{\vec \theta}} \hat H^2 \ket{\psi_{\vec \theta}}}{\braket{\psi_{\vec \theta}|\psi_{\vec \theta}}}-E^2$ denote the mean energy and energy fluctuations in the initial state, respectively.
Thus, the short-time evolution requires only to be capable of measuring the cumulants of energy in the initial condition, which makes this a straightforward solution of the dynamics with NQS.
When considering initial product states, this expansion can be further refined in such a way as to analytically construct an ANN to provide the short-time NQS dynamics by means of classical networks~\cite{Schmitt2018,Verdel2021}, see also Sec.~\ref{subsec:classical_networks}.

While such Taylor expansions are, in principle, already quite powerful, it is finally the main goal of utilizing NQS for real-time evolution to get beyond time scales, which can be captured by a short-time expansion and therefore by a low-order polynomial in $t$.
In the remainder we will introduce and discuss suited methods to achieve this aim.

\subsection{Time-dependent variational principle}
\label{subsec:tdvp}

For practical purposes, it is essential to develop algorithmic means to find the neural network representation of the desired quantum state. In the interest of clarity, we will focus on unitary dynamics of pure states, where the time evolution of the wave function is prescribed by the Schrödinger equation
\begin{align}
    i\frac{d}{dt}\ket{\psi(t)}=\hat H\ket{\psi(t)}\ .
    \label{eq:seq}
\end{align}
\begin{wrapfigure}{r}{0.45\textwidth}
\vspace{-0.8cm}
\center
\includegraphics[width=.45\textwidth]{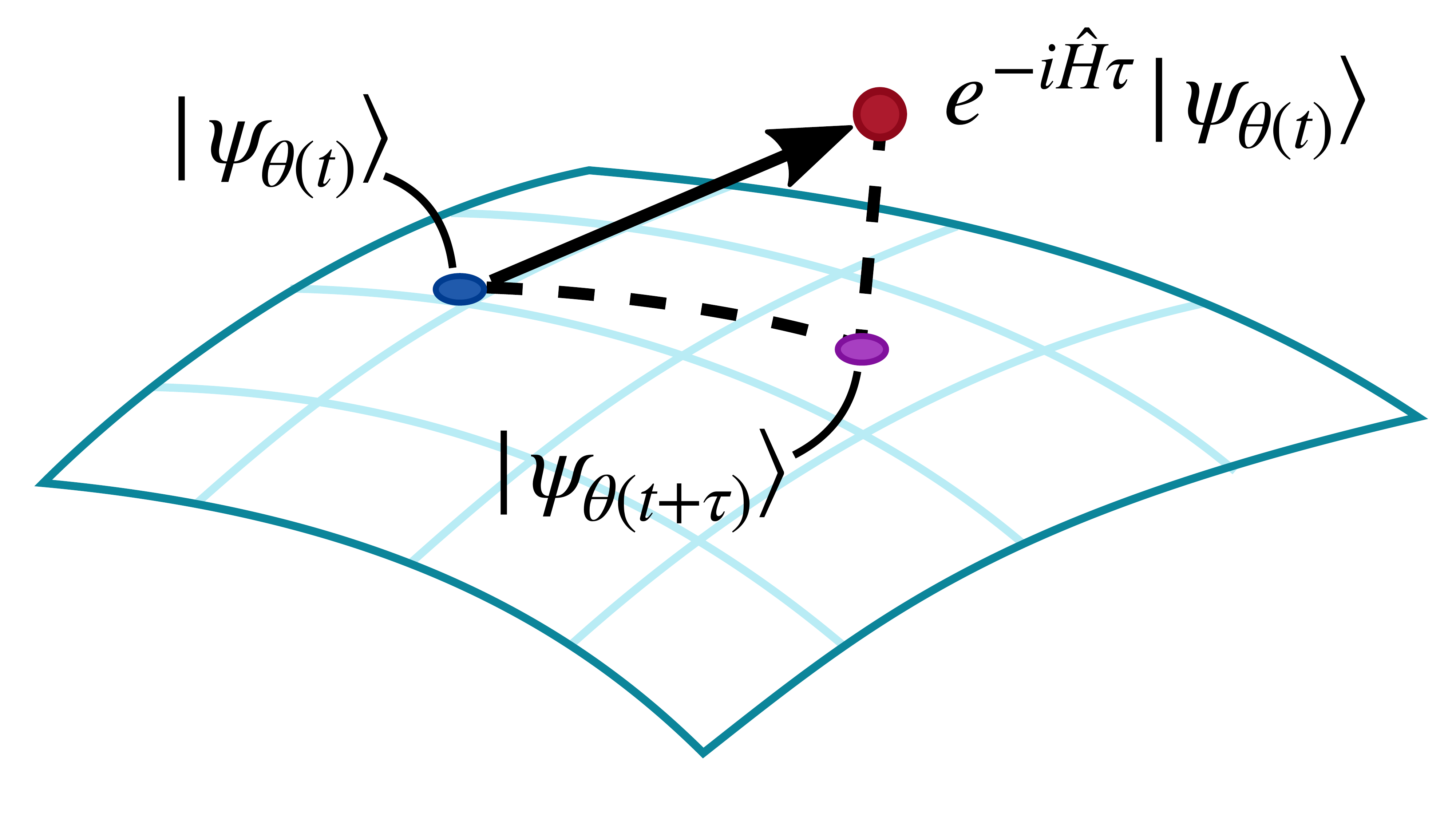}
\caption{Schematic depiction of one step of a time-dependent variational principle. The wave function is propagated by a discrete time step $\tau$ by finding $\vec\theta(t+\tau)$ within the variational manifold (blue), such that the exact evolution $e^{-i\hat H\tau}\ket{\psi_{\vec\theta(t)}}$ is best matched.}
\label{fig:tdvp}
\vspace{-.8cm}
\end{wrapfigure}
The approach discussed in the following can be swiftly generalized to other descriptions of the dynamics and we will comment on that as we go.

Approximating a solution of Schrö\-din\-ger's equation \eqref{eq:seq} means to determine the trajectory $\vec\theta(t)$ in parameter space, such that the corresponding states $\ket{\psi_{\vec\theta(t)}}$ capture the exact time evolution $\ket{\psi(t)}$ most accurately.
In fact, it is often sufficient to be less demanding and to approximate the solution up to normalization and global phase.
This optimization problem to find the time-dependent NQS constitutes a time-dependent variational principle (TDVP).
In the methods described in the following, it amounts to minimizing the distance to the exact solution within the variational manifold of the NQS.
In fact, these methods are discrete time propagation prescriptions.
Given the variational wave function at some time $t$, $\ket{\psi_{\vec\theta(t)}}$, propagating it for a discrete time step $\tau$ means optimizing the time-evolved parameters $\vec\theta(t+\tau)$, such that the exact evolution $e^{-i\hat H\tau}\ket{\psi_{\vec\theta(t)}}$ is best matched, see Fig.~\ref{fig:tdvp}.
This constitutes a projection onto the variational manifold (or its tangent space in an infinitesimal formulation).

In our presentation, we will distinguish two types of algorithms. We will first discuss infinitesimal approaches to optimize locally, where the smallness of the time step $\tau$ is exploited and a continuous evolution of $\vec\theta(t)$ is assumed. After that, we will turn to global approaches, where the full non-convex optimization problem is solved for every time step without further assumptions.

\subsubsection{Infinitesimal approach to local optimization}
\label{subsubsec:infinitesimal}
The anticipated infinitesimal approaches assume that the optimal solution $\vec\theta(t)$ permits a short time expansion at all times of interest of the following form:
\begin{align}
    \ket{\psi_{\vec\theta(t+\tau)}}=\ket{\psi_{\vec\theta(t)}}+\tau\sum_k\dot{\theta}_k\ket{\partial_k\psi_{\vec\theta}(t)}\ ,
    \label{eq:short_time_expansion}
\end{align}
where $\dot{\vec\theta}\equiv\frac{d}{dt}\vec\theta$ and $\partial_k\equiv\frac{\partial}{\partial\theta_k}$. On this basis, there are two different starting points in order to derive a prescription for the time-dependent variational principle: (i) the minimization of a measure of the difference between the exactly evolved state $e^{-i\hat H\tau}\ket{\psi_{\vec\theta(t)}}$ and the variational state $\ket{\psi_{\vec\theta(t+\tau)}}$, and (ii) a stationary action principle. We will see in both cases how the generally non-convex problem of finding the variational update of the parameters at each point in time is rendered convex by employing the short-time expansion \eqref{eq:short_time_expansion} and thereby restricting the search to the local vicinity of the current parameters $\vec\theta(t)$.

\paragraph{Distance minimization}
When aiming to minimize the difference between\linebreak $e^{-i\hat H\tau}\ket{\psi_{\vec\theta(t)}}$ and $\ket{\psi_{\vec\theta(t+\tau)}}$, the first step is to fix a suited metric to quantify the difference. A natural choice is the Fubini-Study metric, 
\begin{align}
    \mathcal D_{FS}(\ket{\psi},\ket{\phi})=\text{arccos}\sqrt\frac{\braket{\psi|\phi}\braket{\phi|\psi}}{\braket{\phi|\phi}\braket{\psi|\psi}}\ ,
    \label{eq:fubini_study}
\end{align}
which is insensitive to changes of the normalization and global phase factors.
With the chosen metric, we define a cost function
\begin{align}
    \mathcal C(\dot\theta)=\mathcal D_{FS}\big(e^{-i\hat H\tau}\ket{\psi_{\vec\theta(t)}},\ket{\psi_{\vec\theta(t)+\tau\dot{\vec\theta}}}\big)\ ,
    \label{eq:cost_fun_inf}
\end{align}
whose minimum defines the optimal solution $\dot{\vec\theta}$. At this point, the smallness of the discrete time step $\tau$ can be utilized in order to render the optimization problem better tractable. A Taylor expansion in the time step $\tau$ yields
\begin{align}
    \mathcal C(\dot{\vec\theta})
    &=
    \tau^2\Big[
    \Big(\sum_k\dot{\theta}_k\bra{\partial_k\psi_{\vec\theta}}-i \bra{\psi_\theta}\hat H\Big)
    \hat G_{\psi_{\vec\theta}}
    \Big(\sum_{k}\dot{\theta}_{k}\ket{\partial_{k}\psi_{\vec\theta}}+i \hat H\ket{\psi_{\vec\theta}}\Big)
    \Big]
    +\mathcal O(\tau^3)\ .
    \label{eq:tdvp_cost}
\end{align}
Here, we introduced the metric tensor of the Fubini-Study metric,
\begin{align}
    \hat G_\psi=\frac{1}{\braket{\psi|\psi}}\mathds{1}-\frac{1}{\braket{\psi|\psi}^2}\ket{\psi}\bra{\psi}\ ,
\end{align}
which is defined by the relation $\mathcal D_{FS}(\ket{\psi},\ket{\psi}+\ket{\delta\psi})=\braket{\delta\psi|\hat G_\psi|\delta\psi}+\mathcal O(\delta^3)$. 
From the geometric perspective, minimizing Eq.~\eqref{eq:tdvp_cost} is equivalent to solving the projected Schrödinger equation $\hat P_{\psi_{\vec\theta}}\big(\frac{d}{dt}+i\hat H\big)\ket{\psi_{\vec\theta}}=0$ with $\hat P_{\psi_{\vec\theta}}$ the tangent space projector~\cite{Hackl2020}.
Expanding the second order term in Eq.~\eqref{eq:tdvp_cost} leads to
\begin{align}
    \mathcal C(\dot\theta)
    =
    \tau^2\Big[
    \sum_{k,k'}S_{k,k'}^{\vec\theta}\dot\theta_k\dot\theta_{k'}
    -2\sum_k\text{Re}\big[F_k^{\vec\theta}\big]\dot\theta_k+\text{Var}_{\psi_{\vec\theta}}(\hat H)
    \Big]
    +\mathcal O(\tau^3)\ ,
    \label{eq:fs_expansion}
\end{align}
where $S_{k,k'}^{\vec\theta}$ is the \emph{quantum geometric tensor} (QGT) of the variational manifold, that is induced by the Fubini-Study metric,
\begin{align}
    S_{k,k'}^{\vec\theta}=\frac{\braket{\partial_k\psi_{\vec\theta}|\partial_{k'}\psi_{\vec\theta}}}{\braket{\psi_{\vec\theta}|\psi_{\vec\theta}}}-
    \frac{\braket{\partial_k\psi_{\vec\theta}|\psi_{\vec\theta}}\braket{\psi_{\vec\theta}|\partial_{k'}\psi_{\vec\theta}}}{\braket{\psi_{\vec\theta}|\psi_{\vec\theta}}^2}\ ,
    \label{eq:tdvp_qgt}
\end{align}
and we introduced the \emph{force vector}
\begin{align}
    F_k^{\vec\theta}=-i\Bigg[\frac{\braket{\partial_k\psi_{\vec\theta}|\hat H|\psi_{\vec\theta}}}{\braket{\psi_{\vec\theta}|\psi_{\vec\theta}}}-
    \frac{\braket{\partial_k\psi_{\vec\theta}|\psi_{\vec\theta}}\braket{\psi_{\vec\theta}|\hat H|\psi_{\vec\theta}}}{\braket{\psi_{\vec\theta}|\psi_{\vec\theta}}^2}\Bigg]\ .
    \label{eq:tdvp_force}
\end{align}
The third term appearing in the second-order contribution to $\mathcal C(\dot{\vec\theta})$ is the energy variance $\text{Var}_{\psi_\theta}\big(\hat H\big)=\frac{\braket{\psi_{\vec\theta}|\hat H^2|\psi_{\vec\theta}}}{\braket{\psi_{\vec\theta}|\psi_{\vec\theta}}}-\frac{\braket{\psi_{\vec\theta}|\hat H|\psi_{\vec\theta}}^2}{\braket{\psi_{\vec\theta}|\psi_{\vec\theta}}^2}$. Through the short-time expansion, the optimization problem becomes convex and we can solve for the stationary point.
This yields the TDVP equation
\begin{align}
    \sum_{k'}\text{Re}\big[S_{k,k'}^{\vec\theta}\big]\dot\theta_{k'}=\text{Re}\big[F_k^{\vec\theta}]\ ,
    \label{eq:tdvp_1}
\end{align}
which is a first order differential equation for the set of variational parameters $\vec\theta$. Notice, however, that the dimensional reduction compared to the original Schrödinger equation comes at the cost of non-linearity of the TDVP equation, because $S_{k,k'}^{\vec\theta}$ and $F_k^{\vec\theta}$ depend non-linearly on $\vec\theta$ in general.

Similarly to physical observables, the QGT and the force vector defined in Eqs.~\eqref{eq:tdvp_qgt} and \eqref{eq:tdvp_force} can be estimated using Monte Carlo sampling, as we will discuss below.

\paragraph{Stationary action principle}
An alternative starting point is a stationary action principle. Schrödinger's equation \eqref{eq:seq} can be obtained by determining a stationary point of the action $\mathcal S=\int dt \mathcal L_{\text{SE}}(\ket{\psi})$ with the Lagrangian $\mathcal L_{\text{SE}}(\ket{\psi})=\braket{\psi|i\frac{d}{dt}-\hat H|\psi}$. The alternative Lagrangian 
\begin{align}
    \mathcal L(\ket{\psi})=\frac{i}{2}\Bigg(\frac{\braket{\psi|\dot\psi}}{\braket{\psi|\psi}}-\frac{\braket{\dot\psi|\psi}}{\braket{\psi|\psi}}\Bigg)-\frac{\braket{\psi|\hat H|\psi}}{\braket{\psi|\psi}}
\end{align}
yields dynamics equivalent to solutions of Schrödinger's equation up to the global phase and normalization \cite{Kramer1981}. For a variational ansatz $\ket{\psi_{\vec\theta}}$, this provides the grounds to formulate a TDVP by demanding stationarity of the action $\mathcal S[\vec\theta(t)]=\int dt \mathcal L(\ket{\psi_{\vec\theta}})$ under variation of the parameters $\vec\theta$, i.e. $\delta \mathcal S[\vec\theta(t)]=0$, resulting in equations of motion
\begin{align}
    \sum_{k'}\text{Im}\big[S_{k,k'}^{\vec\theta}\big]\dot\theta_{k'}=\text{Im}\big[F_k^{\vec\theta}]\ .
    \label{eq:tdvp_2}
\end{align}
Notably, these equations of motion correspond to a classical Hamiltonian dynamics with Hamilton function $\mathcal H(\vec\theta)=\frac{\braket{\psi_{\vec\theta}|\hat H|\psi_{\vec\theta}}}{\braket{\psi_{\vec\theta}|\psi_{\vec\theta}}}$, meaning, that solutions of this TDVP conserve energy by construction \cite{Kramer1981}. 
Let us, moreover, add the following remarks:
\begin{itemize}
\item While Eq.~\eqref{eq:tdvp_1} means solving the projected Schrödinger equation\linebreak $\hat P_{\psi_{\vec\theta}}\big(\frac{d}{dt}+i\hat H\big)\ket{\psi_{\vec\theta}}=0$, Eq.~\eqref{eq:tdvp_2} corresponds to the alternative formulation $\hat P_{\psi_{\vec\theta}}\big(i\frac{d}{dt}-\hat H\big)\ket{\psi_{\vec\theta}}=0$ \cite{Hackl2020}.
Both TDVP quations are equivalent for Kähler manifolds \cite{Brockhoeve1988}, i.e., when the chosen ansatz $\psi_{\vec\theta}(\vec s)$ is a holomorphic function of complex parameters $\theta_k\in\mathbb C$. But in general they yield different solutions \cite{Kucar1987}.
\item Upon solving for the evolution of the complex parameters, the corresponding change of the global phase and normalization can be obtained by plugging the variational state endowed with a complex prefactor, $e^{\varphi}\ket{\psi_{\vec\theta}}$, into Schrödinger's equation. This yields
\begin{align}
    \dot\varphi=-i\frac{\braket{\psi_{\vec\theta}|\hat H|\psi_{\vec\theta}}}{\braket{\psi_{\vec\theta}|\psi_{\vec\theta}}}-\sum_k\dot\theta_k\frac{\braket{\psi_{\vec\theta}|\partial_{k}\psi_{\vec\theta}}}{\braket{\psi_{\vec\theta}|\psi_{\vec\theta}}}\ .
\end{align}
\item The expansion of the Fubini-Study metric in Eq.~\eqref{eq:fs_expansion} provides a measure for the accuracy of the variational solution found in every discrete time step. This quantity is immediately accessible in numerical implementations and it can be used to determine self-consistently, whether the chosen ansatz is expressive enough and whether the dynamics has been accurately captured by the NQS.
\end{itemize}

\paragraph{Monte Carlo estimation} Clearly, practical implementations of the \linebreak TDVP for NQS hinge on the ability to efficiently compute the QGT $S_{k,k'}^{\vec\theta}$ and the force vector $F_k^{\vec\theta}$. For this purpose, it is common practice -- in analogy to the computation of physical operator expectation values discussed in Section \ref{subsubsec:mc_estimation} -- to rewrite all appearing inner products in the form of an expectation value with respect to the Born distribution, in order to facilitate Monte Carlo estimation of the respective quantities. The resulting method for numerical time evolution is called \emph{time-dependent variational Monte Carlo (t-VMC)} \cite{Carleo2012,Carleo2014}. It has been applied successfully in combination with NQS in various applications, some of which beyond the feasibility of other numerical approaches, which we will discuss in Section \ref{sec:applications}. 
Notice, however, that this required numerous methodical tweaks, in order to deal with typical instabilities. We will discuss the necessity of careful regularization as well as the fact, that t-VMC in its current formulation can become ill-defined among the open problems in Section \ref{subsubsec:open_questions_infintesimal}.

\paragraph{Generalization to other first order linear ODEs}
Finally, the approach outlined above for the example of Schrödinger's equation, can be generalized to other linear ODEs of the general form
\begin{align}
    \frac{d\vec y}{dt}=\hat{\mathcal G}\vec y\ .
    \label{eq:general_ode}
\end{align}
In an infinitesimal formulation, the objective is to minimize the length of the difference vector $d\vec y-\hat{\mathcal G}\vec ydt$ as measured by a chosen metric $\mathcal D(\vec y,\vec z)$ with metric tensor $g_{ij}$,
\begin{align}
    ds^2=g_{ij}(d\bar{\vec y}-\hat{\mathcal G}^\dagger \bar{\vec y}dt)_i(d\vec y-\hat{\mathcal G}\vec ydt)_j\ .
\end{align}
Here, we use Einstein summation convention and $\bar{\cdot}$ denotes complex conjugation. Assuming a parametrization $\vec y(t)\equiv \vec y_{\vec\theta(t)}$ yields
\begin{align}
    ds^2=dt^2\Big(
    \underbrace{
    g_{ij}\frac{\partial\bar y_i}{\partial\theta_\mu}\frac{\partial y_j}{\partial\theta_\nu}}_{\equiv S_{\mu\nu}^{\vec\theta}}
    \dot\theta_\mu\dot\theta_\nu
    -2\text{Re}\Big[
    \underbrace{
    \frac{\partial\bar y_i}{\partial\theta_\mu}g_{ij}\big(\hat{\mathcal G}\vec y\big)_j
    }_{\equiv F_\mu^{\vec\theta}}
    \Big]\dot\theta_\mu
    +g_{ij}(\hat{\mathcal G}^\dagger \bar{\vec y}dt)_i(\hat{\mathcal G}\vec ydt)_j
    \Big)
\end{align}
where as above we can identify the induced geometric tensor $S_{\mu\nu}^{\vec\theta}$ and define a force $F_\mu^{\vec\theta}$. Accordingly, demanding a minimal difference at each point in time leads to a TDVP equation of the same form as Eq.~\eqref{eq:tdvp_1}.
An example in the context of quantum mechanics, where the outlined generalization has been used, is the variational solution of Lindblad equations for open system dynamics. Different representations of the density matrix admit a respective corresponding version of the TDVP equation \cite{Hartmann2019, Reh2021}. Beyond the quantum realm, the utility of the approach has been demonstrated for solving high-dimensional Fokker-Planck equations \cite{Reh2022}.

\subsubsection{Global solution}
\label{subsubsec:global}
By releasing the constraint of a sufficiently smooth evolution of the variational parameters, the variational objective can be formulated analogously to Eq.~\eqref{eq:cost_fun_inf} in terms of a cost function
\begin{align}
    \mathcal C(\tilde{\vec\theta})=\mathcal D\big(e^{-i\hat H\tau}\ket{\psi_{\vec\theta(t)}},\ket{\psi_{\tilde{\vec\theta}}}\big)\ ,
\end{align}
where $\mathcal D$ is a suited distance measure of choice and the goal is to find the updated variational parameters $\tilde{\vec\theta}\equiv\vec\theta(t+\tau)$ minimizing it. In order to render the problem tractable, the exact time evolution operator $\hat\Pi_\tau[\hat H]\equiv e^{-i\hat H\tau}$ has to be approximated by a suited numerical approximation $\hat\Phi_\tau[\hat H]$, such that
\begin{align}
    \mathcal C(\tilde{\vec\theta})=\mathcal D\big(\hat\Phi_\tau[\hat H]\ket{\psi_{\vec\theta(t)}},\ket{\psi_{\tilde{\vec\theta}}}\big)\ ,
    \label{eq:cost_fun_glob}
\end{align}
and it's gradient $\nabla_{\tilde{\vec\theta}}\mathcal C(\tilde{\vec\theta})$
can be evaluated efficiently. Thereby, the problem is formulated in close analogy to typical machine learning setups and it becomes amenable to the corresponding sophisticated toolbox for gradient-based optimization in high-dimensional non-convex cost landscapes \cite{Sun2020}. Notice, that the approach is in fact not limited to solving the Schrödinger equation. For example, we can equivalently think of the propagator $\hat\Pi_\tau[\hat H]$ as (the approximation of) a unitary quantum gate or some projection operator, when simulating the effect of projective measurement~\cite{Sinibaldi2023}.

\paragraph{Approximating the time propagator}
Clearly, the success of this approach hinges on a suited choice of the distance measure $\mathcal D$ and the numerical propagator $\hat\Phi_\tau[\hat H]$ and different possibilities have been explored in recent work. Concerning the propagator, it is crucial, that coefficients of the propagated wave function $\braket{\vec s|\hat\Phi_\tau[\hat H]|\psi_\theta}$ can be computed efficiently. In order to achieve this, three types of approaches have been proposed, namely (i) suited Taylor-series expansions, (ii) full generator product expansions, and (iii) decomposition into products of $K$-local operators. Following the first strategy, the implicit midpoint rule, which guarantees unitarity of the propagator, has been employed in Ref.~\cite{Gutierrez2022} and a general framework for diagonally implicit Runge-Kutta schemes has been worked out in Ref.~\cite{Donatella2023}. However, higher order Runge-Kutta schemes require the costly evaluation of matrix elements of powers of the Hamiltonian. Therefore, Taylor root expansion or Pade product expansions, which involve only the generator itself, have been put forward as an effective alternative of type (ii) \cite{Nys2024,Gravina2024}.
Splitting the propagator into a product of $K$-local unitaries, $\hat\Phi_\tau=\prod_l\hat U_\tau^{(l)}$, is most natural, when considering quantum circuits with local gates \cite{Jonsson2018, Medvidovic2021}. For local Hamiltonians, approximate factorizations of the propagator can always be obtained through Suzuki-Trotter decomposition, as considered in this context in Refs.~\cite{Sinibaldi2023,Zhang2024}. Then, the time propagation can be performed by optimizing the parameters for the individual local unitaries sequentially.

\paragraph{The distance measure}
A natural choice for the distance measure would be the Fubini-Study metric \eqref{eq:fubini_study}. Recently, a series of works \cite{Medvidovic2021,Sinibaldi2023} has considered an alternative route via the closely related infidelity
\begin{align}
    \mathcal I(\ket{\psi},\ket{\phi})=1-\frac{\braket{\psi|\phi}\braket{\phi|\psi}}{\braket{\psi|\psi}\braket{\phi|\phi}}\equiv1-\mathcal F(\ket{\psi},\ket{\phi})
\end{align}
or the negative logarithmic fidelity $\mathcal D(\ket{\psi},\ket{\phi})=-\log\big(\mathcal F(\ket{\psi},\ket{\phi})\big)$ \cite{Jonsson2018,Gui2024}. For clarity, we restrict our discussion to the infidelity; the treatment of the logarithmic fidelity follows analogously. In order to facilitate the Monte Carlo estimation necessary for scalability to large system sizes, the corresponding cost function has to be rewritten into the general form
\begin{align}
    \mathcal C(\tilde{\vec\theta})
    =
    \mathcal I\big(\hat\Phi_\tau[\hat H]\ket{\psi_{\vec\theta(t)}},\ket{\psi_{\tilde{\vec\theta}}}\big)
    =
    \sum_{\vec x,\vec y}p(\vec x,\vec y)\mathcal I_{\tilde{\vec\theta}}^{\text{loc}}(\vec x,\vec y)\ ,
        \label{eq:infidelity_cost}
\end{align}
where $p(\vec x,\vec y)$ is a probability over the pairs of computational basis states and $\mathcal I_{\tilde{\vec\theta}}^{\text{loc}}(\vec x,\vec y)$ is the corresponding local infidelity estimator. The expression of the gradient $\nabla_{\tilde{\vec\theta}}\mathcal C(\tilde{\vec\theta})$ in terms of suited expectation values follows in the usual manner of variational Monte Carlo \cite{Jonsson2018, Sinibaldi2023}. 
The choice of a well-behaved estimator turns out to be crucial and we will discuss this issue further in Section \ref{subsubsec:open_questions_global}.

For one concrete example, consider a unitary $\hat\Phi_\tau[\hat H]$. In that case, it is useful to choose the product of Born probabilities for the joint distribution
\begin{align}
    p(\vec x,\vec y)=\frac{|\psi_{\tilde{\vec\theta}}(\vec x)|^2}{\braket{\psi_{\tilde{\vec\theta}}|\psi_{\tilde{\vec\theta}}}}
    \frac{|\psi_{\vec\theta}(\vec y)|^2}{\braket{\psi_{\vec\theta}|\psi_{\vec\theta}}}
\end{align}
and the local estimator
\begin{align}
    \mathcal I_{\tilde{\vec\theta}}^{\text{loc}}(\vec x,\vec y)=1-\frac{\braket{\vec x|\hat\Phi_\tau[\hat H]|\psi_{\vec\theta}}}{\braket{\vec x|\psi_{\tilde{\vec\theta}}}}
    \frac{\braket{\vec y|\hat\Phi_\tau^\dagger[\hat H]|\psi_{\tilde{\vec\theta}}}}{\braket{\vec y|\psi_{\vec\theta}}}\ .
    \label{eq:local_infidelity}
\end{align}
Thereby, the expectation value can be factorized and more costly sampling from propagated wave functions $\hat\Phi_\tau[\hat H]\ket{\psi_{\vec\theta}}$ can be avoided, which is necessary in other formulations. 

Various alternative distance measures have been considered in the literature, such as different variants of least squares cost \cite{Kochkov2018,Westerhout2020,Gutierrez2022} or a combination of Kullback-Leibler divergence for the amplitude part and least squares for the phase of the wave function \cite{Lin2021}.

\paragraph{Generalization to other first-order linear ODEs}
Just as for the local optimization, the generalization of global optimization strategies to other first order linear ODEs is straightforward. For example, Ref.~\cite{Luo2022} implemented an approach based on a second-order forward-backward trapezoid method to solve a Lindblad equation for open system dynamics.

\subsection{Other evolution techniques}
\label{subsec:other_evolution}

\paragraph{Explicit time dependence}
The starting point of the TDVP introduced in Section \ref{subsec:tdvp} is a variational ansatz for the wave function $\ket{\psi_{\vec\theta(t)}}$, where the time dependence is introduced via a time dependence of the variational parameters $\vec\theta(t)$.
In recent work \cite{Walle2024,Sinibaldi_galerkin2024}, an alternative approach has been proposed, namely to consider an explicitly time-dependent wave function with time-independent parametrization,
\begin{align}
    \ket{\psi_\theta(t)}=\sum_{\vec x}\psi_\theta(\vec x, t)\ket{\vec x}\ .
\end{align}
Based on this, an optimized set of parameters to best approximate a solution of Schrödinger's equation on a given time interval $t\in[0,T]$ can be found by minimizing the loss function
\begin{align}
    \mathcal L(\vec\theta)=\frac{1}{T}\int_0^T\Big\lVert\frac{d}{dt}\ket{\psi_{\vec\theta}(t)}+i\hat H\ket{\psi_{\vec\theta}(t)}\Big\rVert^2\ dt\ .
\end{align}
For simplicity of the presentation, we assume a normalized ansatz and the initial condition $\ket{\psi_{\vec\theta}(t=0)}=\ket{\psi_0}$ has to be fixed by constructing the ansatz accordingly. Upon approximating the time integral with a Riemann sum, the value of the loss $\mathcal L(\vec\theta)$ and it's gradient $\nabla_{\vec\theta}\mathcal L(\vec\theta)$ can be formulated as a sum of expectation values with respect to the Born distribution $|\psi_{\vec\theta}(t)|^2$. Accordingly, the Monte Carlo sampling strategies introduced above are applicable. Notice, that the approach is crucially enabled by automatic differentiation techniques, which facilitate dealing with the time derivative, $\frac{d}{dt}\ket{\psi_{\vec\theta}(t)}$, irrespective of the specific functional form.

The time-dependent wave function $\psi_{\vec\theta}(\vec x, t)$ can be constructed in various ways. Walle \textit{et al.} \cite{Walle2024} chose a transformer neural network architecture, which, besides the basis configuration $\vec x$, processes the time $t$ as additional input to compute the wave function coefficient. Sinibaldi \textit{et al.} \cite{Sinibaldi_galerkin2024} composed the ansatz as a time-dependent superposition of static NQS wave functions, where the time-dependent coefficients of the superposition are parametrized in the form of a Fourier series. In proof-of-principle demonstrations, both approaches were shown to achieve state of the art performance in simulating dynamics of large two-dimensional quantum magnets.

\paragraph{Exact application of diagonal operators}
The action of operators, that are diagonal in the computational basis, $\braket{\vec x|\hat O|\vec x'}=O_{\vec x}\delta_{\vec x,\vec x'}$, can be exactly accounted for without the need for variational optimization. Since 
\begin{align}
    \ket{\tilde\psi_{\vec\theta}}\equiv\hat O\ket{\psi_{\vec\theta}}=\sum_{\vec x}\psi_\theta(\vec x)O_{\vec x}\ket{\vec x}\ ,
\end{align}
the new state $\ket{\tilde\psi_{\vec\theta}}$ is represented by the updated variational ansatz $\tilde\psi_{\vec\theta}(\vec x)=O_{\vec x}\psi_{\vec\theta}(\vec x)$. This approach was employed in Ref.~\cite{MendesSantos2023} to compute two-time correlation functions of spin operators $\hat O=\hat S_i^z$. Similarly, it is always possible to choose a part of a unitary gate set diagonal in the computational basis. Going beyond the simple adaptation of the variational ansatz as described above, it was demonstrated in Refs.~\cite{Jonsson2018,Medvidovic2021}, that single qubit rotations $\hat R^z(\phi)$ can always be captured within a Restricted Boltzmann Machine architecture by adapting the visible bias values. And the action of controlled Z-rotations $CR^z(\phi)$ as well as RZZ rotations ($RZZ(\phi)=e^{i\phi \hat\sigma_i\hat\sigma_j}$) corresponds to adding an additional hidden unit with specific couplings connected to the respective qubits. A universal gate set will, however, always contain at least one gate, that is not diagonal in the chosen computational basis.

\paragraph{Dynamics as a ground state problem}
The Feynman-Kitaev construction for time evolution employs an ancillary ``time register'' to represent a time-dependent quantum state $\ket{\psi(t)}$ at discrete time points $t$ in a time-independent form via $\ket{\phi}\propto\sum_t\ket{\psi(t)}\otimes\ket{t}$. Based on this, it is possible to identify the solution $\ket{\phi}$ corresponding to a given unitary time evolution as the ground state of a suitably chosen Hamiltonian $\hat H_{FK}$ \cite{McClean2013}. Therefore, the established machinery of variational Monte Carlo combined with NQS for ground state search can be employed to find such Feynman-Kitaev states. In a proof-of-principle study, Ref.~\cite{VargasCalderon2023} demonstrated the time evolution of a one-dimensional transverse-field Ising model. The increasingly challenging optimization was, however, identified as a main factor limiting the reached system sizes and time scales.

\section{Applications}
\label{sec:applications}

\subsection{Isolated system dynamics}
The non-equilibrium dynamics of isolated many-body systems has received substantial theoretical attention in recent years, motivated in particular by the unprecedented capabilities for experimental realization in quantum simulators \cite{Altman2021}. Various implementations, such as cold atoms in optical lattices \cite{Gross2017}, Rydberg atom arrays \cite{Browaeys2020}, or Josephson junction arrays \cite{Kockum2019} specifically target the frontier of two-dimensional quantum matter, which, however, also represents a significant challenge in the theoretical description.
This serves as a prime motivation to study the unitary dynamics of interacting quantum many-body systems with the NQS approach, which has already seen tremendous progress as will be summarized in the following.

\begin{figure}[t!]
\includegraphics[width=\textwidth]{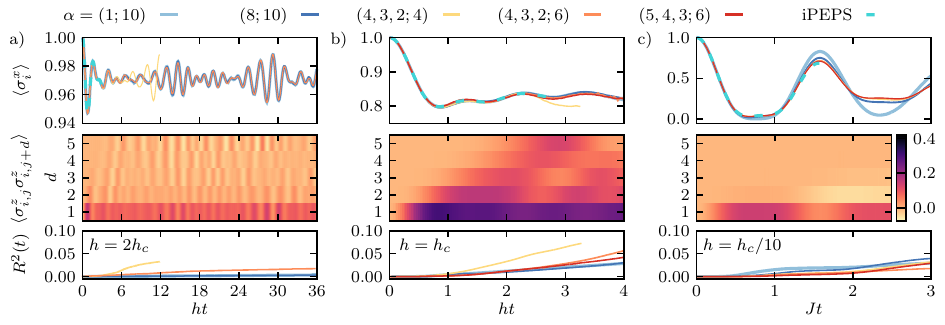}
\caption{Benchmark results for quench dynamics simulated with NQS. The two-dimensional transverse-field Ising magnet is initialized in a paramagnetic state and time-evolved at three different values of the magnetic field ranging across the phase diagram. iPEPS data from Ref.~\cite{Czarnik2019} are included as reference and the NQS approach consistently reaches comparable or longer times. Figure adapted from Ref.~\cite{Schmitt2020}.}
\label{fig:nqs_tdvp_benchmark}
\end{figure}

\paragraph{Proof of principle}
A first proof of principle for time evolution with NQS wave functions was given by Carleo and Troyer in their seminal work \cite{Carleo2017}. They demonstrated, that time evolution reaching short to intermediate times is possible using time-dependent variational Monte Carlo based on the TDVP as described in Section \ref{subsubsec:infinitesimal} for quenches in two one-dimensional quantum magnets, namely the transverse-field Ising model and the antiferromagnetic Heisenberg model. However, it was found in a subsequent work, that in tilted field Ising models (including the integrable transverse-field case) quenches to the vicinity of the critical point appear very challenging to simulate already at short times and small system sizes \cite{Czischek2018}.
By contrast, it has been shown in Ref.~\cite{Schmitt2020} for the example of quantum Ising models, that NQS-based simulations are competitive with and partly already superior to other state-of-the-art methods when it comes to far-from-equilibrium dynamics in two spatial dimensions, see Fig.~\ref{fig:nqs_tdvp_benchmark}. 
The key to reaching the presented time scales was a careful regularization of the pseudo-inverse of the QGT, a suited adaptive time step integrator, and the choice of non-singular activation functions for the complex parametrized neural networks. Remarkably, the convergence with network size was very rapid for most of the results, indicating that network expressivity is not the major limiting factor for the considered situations. Most recently, the same benchmark problem has been studied using a global optimization approach as described in Section~\ref{subsubsec:global}, reaching a comparable time scale for one of the physical parameter sets \cite{Gravina2024}, and the achieved timescales have been extended using the infinitesimal TDVP with a ResNet architecture \cite{Chen2025}.
Moreover, the applicability of the TDVP has been demonstrated to time-evolve initially thermal states upon purification in the form of so-called neural thermofields \cite{Nys2024-2}.

\paragraph{Phase transition dynamics} For systems driven across a continuous phase transition, the Kibble-Zurek mechanism predicts, that the resulting excitations are fully characterized by the universal properties of the underlying transition \cite{Kibble1980,Zurek1985}. Such dynamical universality has been explored extensively in one-dimensional quantum lattice models. However, numerical investigations in interacting two-dimensional systems have long remained elusive. By combining the complementary strengths of state-of-the-art tensor network and NQS methods, Ref.~\cite{Schmitt2022} reported the first numerical observation of universal dynamical behavior for an interacting two-dimensional quantum magnet, including the full scaling form of a correlation function. For this purpose, it was crucial, that NQS rendered system sizes up to $20\times20$ spins tractable. The same problem has been addressed in Ref.~\cite{Donatella2023} with a global optimization approach for a range of smaller system sizes.
Phase transition dynamics can nowadays also be realized in quantum simulators and the numerical findings are in line with recent experiments using Rydberg atom arrays \cite{Ebadi2021}. More than that, NQS have been identified as a promising candidate to classically simulate the dynamics of detailed models of Rydberg atom arrays. In Ref.~\cite{MendesSantos2024}, the authors proposed to analyze many-body quantum states in terms of so-called wave function networks, which were shown for both experimental data and NQS-based simulations to yield a scale-free structure for a ramp across a quantum phase transition.
Similarly, NQS were considered among the numerical simulation techniques used to gauge the classical simulability of dynamics realized on a quantum annealer \cite{King2024}.

\paragraph{Dynamics of fundamental excitations in two-dimensional magnets} 
The dynamics of magnonic excitations is both of fundamental interest and relevant for technological applications such as high-speed low-energy data processing. Femtosecond x-ray techniques enable the probing of magnon dynamics in real materials \cite{Buzzi2018,Mazzone2021}. When studying the response of a two-dimensional Heisenberg antiferromagnet to a short perturbation of the exchange interaction with NQS-based simulations, Ref.~\cite{Fabiani2021} revealed an unexpectedly high magnon velocity on short length and time scales. This observation was attributed to strong magnon-magnon interactions. Furthermore, NQS were employed to investigate to what extent predictions from linear spin wave theory about the evolution of entanglement upon quenching the exchange interaction persist \cite{Fabiani2022}.
Another intriguing magnetic structure in two spatial dimensions is the skyrmion. Understanding their dynamical behavior will be crucial for their controlled manipulation in envisioned skyrmion-based storage devices. In Ref.~\cite{Joshi2024}, the dynamics of quantum skyrmions was investigated based on NQS simulations, demonstrating for example their decay, when multiple skyrmions interact with each other.

\paragraph{Spectral functions} The NQS techniques for real time evolution lend themselves as a tool to probe spectral properties in the low-energy regime. The evolution following small time-dependent perturbations of Heisenberg antiferromagnets was used in Ref.~\cite{Fabiani2019} to obtain the frequency-resolved spin structure factor, which is experimentally accessible via inelastic x-ray scattering. In a similar spirit, Ref.~\cite{MendesSantos2023} obtained the spin structure factor in two-dimensional quantum Ising models, however based on directly incorporating the action of excitation operators in the NQS ansatz as described in Section \ref{subsec:other_evolution}. Remarkably, the approach accurately captured the diverging time scales close to the quantum phase transition up to system sizes of $24\times24$ spins.

\paragraph{Continuous variables: Bosons and Fermions}
While the first works considered spin systems, the methodology has in the meantime also been extended to continuous variable systems. On the one hand, Ref.~\cite{Medvidovic2023} demonstrated accurate simulations for quenches in Josephson junction arrays of sizes up to $8\times8$ sites, including the approach of observables to their thermal equilibrium value. On the other hand, Ref.~\cite{Nys2024} reports results for the time evolution of many-electron systems in different non-equilibrium situations using position space wave functions with a neural backflow architecture implementing the proper anti-symmetrization for the fermionic particles.

\paragraph{Many-body quantum chaos}
Out-of-time-order correlators (OTOCs) have been established over the past decade as a probe of information scrambling and chaos in many-body quantum systems \cite{Hosur2016}. Ref.~\cite{Wu2020} proposes an approach to access OTOCs in NQS simulations based on the overlap of two time-evolved wave functions, reporting results for the two-dimensional quantum Ising model with up to 100 spins and up to times of one unit of the inverse Ising coupling.

\paragraph{Renormalization group approach for disordered systems}
Renormalization group (RG) approaches are of key importance for the understanding of phases and phase transitions in matter.
When it comes to quantitatively describing the dynamics of quantum many-body systems by means of RG approaches, it becomes relevant to not only take into account the final renormalized Hamiltonian, but also the whole renormalization group transformation.
In Ref.~\cite{Burau2021} a scheme has been presented where the action of the renormalization group transformation onto a quantum state was accounted for by means of NQS through interpreting this transformation as an effective Hamiltonian time evolution.
This has allowed to describe the nonequilibrium dynamics of a many-body localized quantum spin model both in 1D and 2D for large system sizes and times thereby monitoring for the first time the dynamical buildup of many-body localized spin-glass order.

\subsection{Quantum circuits}
Motivated by the emergence of numerous platforms for universal circuit-based quantum computation, the numerical simulation of quantum circuits has received increasing attention in recent years. In principle, the action of any unitary gate $\hat u$ can be related to a Hamiltonian dynamics described by Schrödinger's equation with a corresponding Hamiltonian $\hat H_u$, 
\begin{align}
    \hat u\ket{\psi}=e^{-i\hat H_ut_u}\ket{\psi}\ .
\end{align}
Accordingly, the techniques described in Sections \ref{subsec:tdvp} are generally applicable to simulate unitary circuits. However, quantum circuits can exhibit additional structure, that can be exploited to simplify the numerical simulations. In particular, the action of gates, which are diagonal in the computational basis, can be analytically accounted for, when updating the NQS, without the need for variational optimization, see Section \ref{subsec:other_evolution}. Ref.~\cite{Jonsson2018} exploited this fact for simulations, where only the action of Hadamard gates had to be approximated in the spirit of Section \ref{subsubsec:global}. Thereby, the authors demonstrate the simulation of Hadamard transforms and truncated Fourier transforms of entangled initial states on arrays of up to 64 qubits. They show that the variational error is equivalent to a noise level of $10^{-3}$ when considering a simple Pauli noise channel. In a subsequent work, similar techniques were used to simulate a Quantum Approximate Optimization Algorithm \cite{Medvidovic2021}.

While analytically tractable gates can reduce the cost of simulating quantum circuits, mid-circuit projective measurements pose a significant challenge. The projection required to obtain the post-measurement state typically introduces strictly vanishing wave function coefficients in the computational basis. Such states can on the one hand be hard to represent for neural networks, and on the other hand the VMC techniques become ill-defined, see Ref.~\cite{Sinibaldi2023} and Section \ref{subsubsec:open_questions_infintesimal}. In Ref.~\cite{Sinibaldi2023} it was, however, shown that these obstacles can be overcome using a global optimization approach. As a showcase application, the authors demonstrated the successful simulation of a measurement-induced phase transition in a two-dimensional array of up to $6\times6$ qubits.

\subsection{Open system dynamics}
\begin{wrapfigure}{r}{0.45\textwidth}
\vspace{-.8cm}
\includegraphics[width=0.45\textwidth]{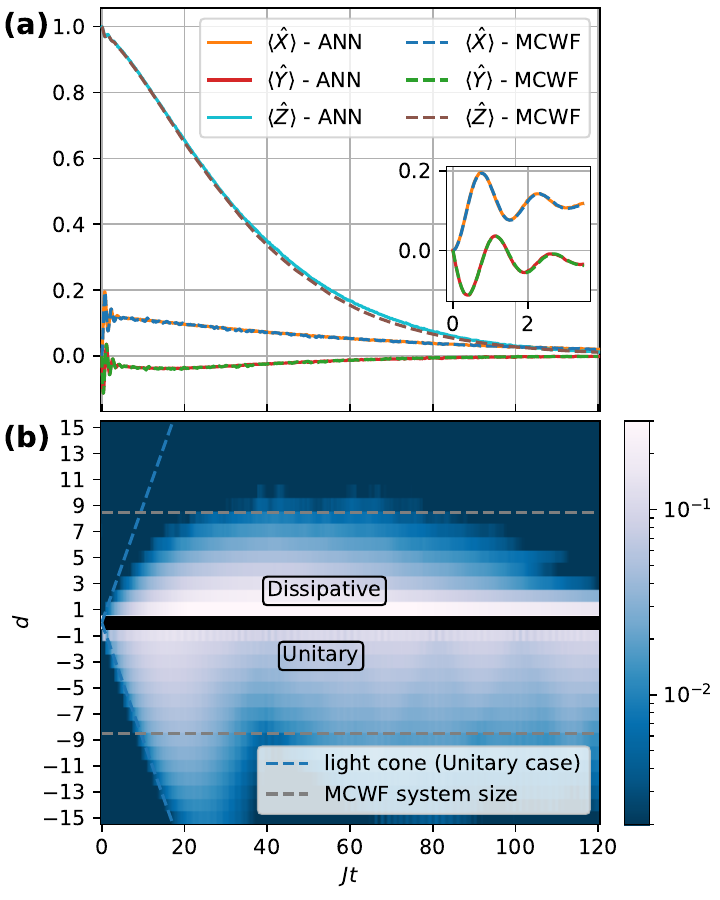}
\vspace{-.7cm}
\caption{Dissipative dynamics of a quantum spin chain. (a) The NQS approach (ANN) accurately captures both the coherent oscillations at short times and the approach to the late time steady state. (b) Impact of dissipation on correlation spreading in comparison with isolated system dynamics. Figure adapted from Ref.~\cite{Reh2021}.}
\label{fig:open}
\vspace{-.5cm}
\end{wrapfigure}
While in current experimental platforms the dynamics of quantum many-body systems can be considered isolated to a high degree of accuracy for large time scales, it is nevertheless important to recognize that the coupling to an environment is inevitable. 
The resulting dissipation can give rise to new phenomena and it may be used as a resource for designing system properties \cite{Diehl2008,Diehl2011,Harrington2022,Khasseh2023}. 
For Markovian dynamics of the bath, open system dynamics can be modeled using a Lindblad equation $\frac{d}{dt}\rho=\mathcal L[\rho]$. Simulating such dynamics is computationally often even more challenging than isolated system dynamics, because in this case a full mixed density matrix has to be accounted for. Similar to the pure states discussed before, ANNs may be used for compressed representations of the density matrix. Given an ANN ansatz, the Lindblad equation can be solved using the techniques described in Section \ref{subsec:tdvp}. Notice, however, that physical density matrices have to obey additional constraints, namely hermiticity and positive-semidefiniteness. 
The focus of pioneering works until now was to benchmark different types of density matrix representations and time evolution techniques. In a first work, Hartmann and Carleo encoded the density matrix in a Restricted Boltzmann Machine, that was constructed to obey hermiticity \cite{Hartmann2019}. They demonstrated time evolution using a TDVP equation, but the reached timescale remained rather limited. Subsequently, Reh \emph{et al.} derived a TDVP equation for density matrices represented in terms of the probability distribution of a Positive Operator Valued Measure \cite{Reh2021}. Encoding the probability distribution in a recurrent neural network, they achieved an accurate simulation of the complete approach to a final steady state of a few dozen qubits, see Fig.~\ref{fig:open}. In Ref.~\cite{Luo2022} it was shown, that the same type of ansatz is amenable to time evolution following the global optimization strategy described in Section \ref{subsubsec:global}. While all these approaches didn't guarantee positivity of the encoded density matrix, a strategy to enforce both positivity and hermiticity was introduced in Ref.~\cite{Vicentini2022} and long time evolution was used, in order to obtain steady states.

\subsection{Analytical solution through classical networks}
\label{subsec:classical_networks}

In the full-fledged scenario the NQS approach represents a purely numerical technique to solve for the quantum many-body problem.
It is, however, important to emphasize that also analytical access is possible.
This holds, in particular, in case the dynamical system at hand exhibits a small parameter suitable for performing a perturbative expansion.
This has been formalized in the context of so-called classical networks~\cite{Schmitt2018,Verdel2021}, which have been shown to be exactly mappable to ANNs~\cite{Schmitt2018}.
Often, in leading order they lead to Jastrow-type variational wave functions, which have been already successfully applied to quantum dynamical problems beyond one dimension~\cite{Blass2016,Comparin2022}. 
When going beyond leading order, the classical network approach provides a constructive algorithm to add additional higher-order interactions in a controlled manner.
Notably, such an analytical approach allows to determine both the network structure as well as the exact values of its weights.
This method has been of great success in describing in particular the dynamics of two-dimensional quantum spin models~\cite{Schmitt2018,Verdel2021}, interacting lattice gauge theories~\cite{Karpov2021,Verdel2021}, or many-body localized systems~\cite{Tomasi2019}.

\section{Open questions and challenges}
\label{sec:challenges}

The utilization of NQS for the description of the dynamics of interacting quantum matter has seen tremendous progress in recent years, as summarized in the previous section.
This progress has allowed NQS to push the frontier of computational methods in particular concerning quantum many-body systems in two spatial dimensions.
However, for the future it will be essential to achieve further advances as the NQS approach is still facing limitations.
In the following, it will be the goal to summarize and discuss key open questions and challenges as well as potential routes for their resolution.
Overall, any variational method to solve the quantum many-body problem is built upon two pillars:
\begin{enumerate}
    \item Expressivity: Which quantum states can be captured efficiently
with the variational ansatz and what defines its limitations? Which physical properties delineate the scope of the ansatz?
    \item Optimization: Can we optimize or train the variational ansatz efficiently and how can we make maximal use of its expressivity? Can optimizability ultimately represent a prohibitive restriction?
\end{enumerate}
Clearly, optimal performance is reached when both pillars can operate at maximal capability.
In the following we will discuss the open questions and challenges in the light of these two main pillars.

\subsection{Expressivity}
\label{subsec:challenges_expressivity}

\paragraph{Physical understanding}
In view of the success of ANNs in the field of computer science it is straightforward to recognize that, in principle, ANNs have an immense potential to encode a wide variety of quantum many-body wave functions.
However, so far it has remained unclear what are the general limitations.
This concerns in particular the limited physical understanding of NQS: what physical properties can be represented well and which ones not?
While it has remained a challenge to identify the overall limitations, the literature covers several individual instances where it has been shown that certain physical circumstances don't represent limitations.
This includes quantum entanglement, one of the key limiting factors of tensor network approaches.
It has been shown that there exist highly entangled volume-law states that can be represented by NQSs with a limited number of parameters \cite{Deng2017,Levine2019,Sharir2022}.
Also, long-range interacting models and systems with permutation symmetry can be efficiently captured with a limited amount of required network parameters~\cite{Ballar2024}.

On a more general level, what has been realized recently is that the expressivity of an NQS can depend crucially on the chosen computational basis~\cite{Yang2024}.
Notice, however, that this work doesn't cover the case of complex amplitudes, which is relevant for dynamics.
When it comes to simulation of dynamics with NQS, it has already been recognized that a continuous temporal adaption of the computational basis can be suitable~\cite{MendesSantos2024}.
Overall, it appears likely that any progress in understanding the physical limitations of NQS also has to incorporate a suitable choice of the basis.

\paragraph{Temporal complexity}
It is the general expectation that time-evolved quantum many-body states become more complex for progressing time.
Accordingly, the larger the overall targeted simulation time it is natural to expect that also the ANNs underlying the NQSs have to grow in size.
In this light it might appear rather surprising that the typical network sizes for time-evolved quantum matter, which have been studied so far in the literature, have been relatively small, at least as compared to some ground-state simulations~\cite{AoChen2024}.
These empirical observations of limited required network sizes are also in contrast to an attempt to quantify the temporal complexity growth by means of a supervised learning approach, where numerical evidence suggests a rapid exponential temporal increase of required neural network sizes as a function time, which according to the utilized method grows even stronger than for tensor networks~\cite{Lin2022}.
While the exponential growth by itself might just reflect the overall expected complexity increase to solve quantum dynamics for longer times, the natural interpretation of this work is that the choice of ANN is crucial.
This holds in particular because tensor networks are a subclass of all NQS \cite{Sharir2022} so that upon properly choosing the ANN, at least the parameter scaling of tensor networks has to be achievable.
On a general level, the observation that only limited ANN sizes seem to be relevant in the solution of the dynamics of quantum matter can be attributed to the principles of locality and causality.
Specifically, due to Lieb-Robinson bounds correlations develop only within a light-cone \cite{Nachtergaele2010} so that only networks with limited local connections are required.
This overall suggests, that during dynamics it might be optimal to sequentially adapt the network over time as correlations spread yielding ideally an adaptive growth of the underlying ANN.

\subsection{Optimization}
\label{subsec:challenges_optimization}

\begin{figure}
\center
\includegraphics[width=.8\textwidth]{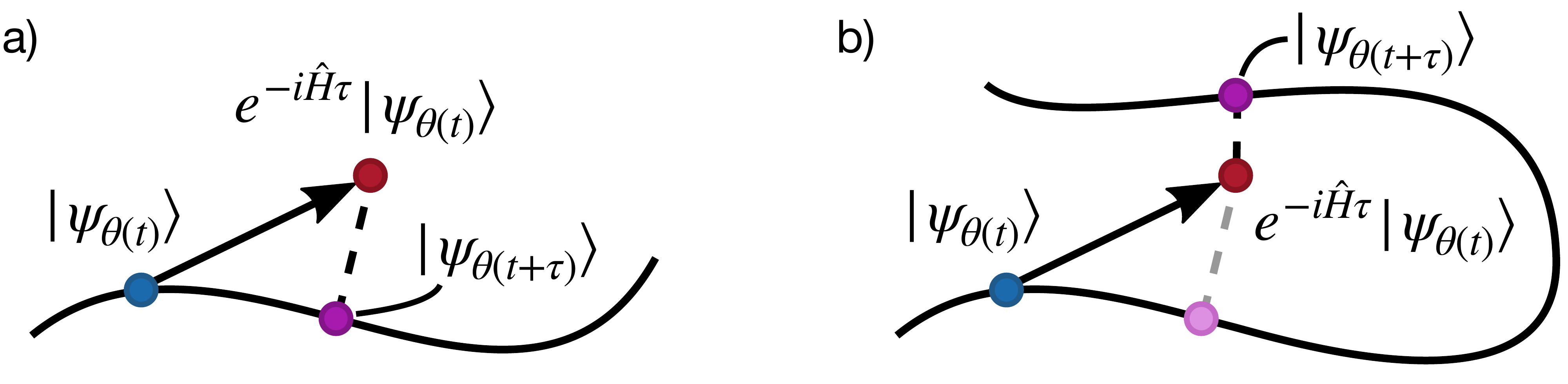}
\caption{Illustration for the key difference between local and global optimization. (a) A case, where the optimal solution $\vec\theta(t+\tau)$ lies in the local vicinity of $\vec\theta(t)$. (b) An example, where the optimal solution corresponds to a large jump in parameter space.}
\label{fig:glob_vs_loc}
\end{figure}
As we outlined above, the state of the art time evolution methods for NQS largely fall into two categories: those solving a TDVP equation for local optimization and those employing gradient based optimization for global optimization. The major difference between both approaches is that the former relies on the existence of a well-defined time-derivative $\dot{\vec\theta}(t)$ for all times $t$, see Eq.~\eqref{eq:short_time_expansion}. This assumption is valid for smooth wave function manifolds and whenever the optimal solution to the TDVP at time $t$ can be found in the local vicinity of $\vec\theta(t)$ as depicted in Fig.~\ref{fig:glob_vs_loc}a. 
While the former has to be ensured when constructing the ansatz, exceptions from the latter are conceivable when the manifold folds within the embedding space as shown in Fig.~\ref{fig:glob_vs_loc}b. In such cases, the optimal solution for a small time step, while close in the full Hilbert space, may lie far away in parameter space.

A common challenge for both global and local approaches to the optimization problem lies in the fact, that the employed tools rely on a number of hyperparameters for sampling, regularization, and optimization. The optimal choice of parameters will depend on the characteristics of the wave function $\ket{\psi_{\vec\theta(t)}}$. At this point, the hyperparameters are typically chosen on a case-by-case basis and often kept fixed during the time evolution. For further progress, it will be key to develop adaptive schemes for automated and time-dependent choice of hyperparameters. 

In the following, we will separately discuss further open questions concerning either of the two approaches.

\subsubsection{Infinitesimal approaches to local optimization}
\label{subsubsec:open_questions_infintesimal}

\paragraph{Ill-defined Monte Carlo estimators}
The estimation of force vectors and quantum geometric tensors by sampling the Born distribution relies in general on non-vanishing wave function amplitudes $|\psi_{\vec\theta}(\vec x)|>0\ \forall\ \vec x$. Formally, if $|\psi_{\vec\theta}(\vec x)|=0$ for some $\vec x$, the resulting estimators acquire unknown biases and become uncontrolled \cite{Sinibaldi2023}. Consider, for concreteness, the QGT -- similar reasoning applies to the force vector.
Formally, in the presence of vanishing amplitudes, the traditionally used estimator takes the form
\begin{align}
    \frac{\braket{\partial_k\psi_{\vec\theta}|\partial_{k'}\psi_{\vec\theta}}}{\braket{\psi_{\vec\theta}|\psi_{\vec\theta}}}= \sum_{\vec x^*: \psi_{\vec\theta}(\vec x^*)=0}\frac{\partial_{k}\bar\psi_{\vec\theta}(\vec x^\ast)\partial_{k'}\psi_{\vec\theta}(\vec x^\ast)}{\braket{\psi_{\vec\theta}|\psi_{\vec\theta}}} + \sum_{\vec x:|\psi_{\vec\theta}(\vec x)|>0} \frac{|\psi_{\vec \theta}(\vec x)|^2}{\braket{\psi_{\vec\theta}|\psi_{\vec\theta}}} \bar\Gamma_k^{\vec\theta}(\vec x)\Gamma_{k'}^{\vec\theta}(\vec x)
    \label{eq:mc_bias}
\end{align}
with $ \Gamma_k^{\vec\theta}(\vec x) = \partial_{k}\psi_{\vec \theta}(\vec x)/\psi_{\vec \theta}(\vec x)$, cf.~Eq.~\eqref{eq:tdvp_qgt}. The contributions with vanishing wave function coefficients cannot be included in the expectation value with respect to the Born distribution. Therefore, the estimator becomes biased. Notice, however, that for typical many-body states, wave function amplitudes as well as their partial derivatives are exponentially small in the system size.
Therefore, the anticipated bias effect will only be relevant when the wave function vanishes on a finite fraction of basis configurations, which, however, is not a generic scenario.
Thus, in the typical cases, the bias term in Eq.~\eqref{eq:mc_bias} only provides an exponentially small correction for large systems which can be safely neglected.

For larger system sizes, the vanishing wave-function amplitudes can generate two different issues.
First, already the presence of individual amplitudes, that are very small compared to the typical magnitude, can lead to high variances, such that the Monte Carlo estimation becomes challenging. 
Small amplitude configurations can contribute very large logarithmic derivatives $\Gamma_{k}^{\vec\theta}(\vec x)$, potentially leading to an intractable variance of the estimator. This means, that excessive Monte Carlo sample sizes are required for reliable estimation, see Eq.~\eqref{eq:MC_error}. 

Second, at large system sizes vanishing wave-function amplitudes can indicate non-analytic temporal behaviors.
Actually, the occurrence of vanishing wave function amplitudes in large systems is in fact generic.
Considering a quantum quench with initial state $\ket{\psi_0}$, the time-evolved wave function amplitude $\psi(\vec x,t)$ can be written as $ \psi(\vec x,t)= \bra{\vec x} e^{-i\hat Ht} \ket{\psi_0}$.
Such objects have been studied extensively in the context of dynamical quantum phase transitions (DQPTs), which occur generically under strong nonequilibrium conditions beyond linear response~\cite{Heyl2013,Heyl2018}.
Importantly, $\psi(\vec x,t)$ can be formally interpreted as a partition function, the vanishing of which is in a one-to-one correspondence with a phase transition (in the thermodynamic limit) and therefore a point of singular behavior as a function of time $t$.
Accordingly, the solution of the TDVP has to become nonanalytic at $t^\ast$, meaning that the corresponding set of nonlinear ODEs has nonanalytic solutions.
Overall, such DQPTs cannot be avoided without a relevant perturbation in the renormalization group sense, as usually the case at equilibrium phase transitions.
Consequently, vanishing wave-function amplitudes for large systems have an underlying physical reason, occur generically for strong nonequilibrium scenarios, and cannot be simply circumvented.
It is an open question, whether these obstacles constitute fundamental limitations of the VMC method, or whether new strategies can be found to circumvent them.
One potential way of dealing with the singular behavior associated with these vanishing wave function amplitudes is to target a (at least piecewise analytic) global solution for instance by means of global optimization approaches~\cite{Sinibaldi2023}.
High variance of the estimators may be mitigated by suitably chosen importance sampling schemes.

\paragraph{Noisy TDVP equation}
It is important to recognize, that t-VMC relies on estimated TDVP equations, which inevitably involve statistical noise.
This implies that the generic scenario in local approaches is to solve a set of noisy non-linear ODEs.
While the analysis of signal-to-noise ratios allows to eliminate particularly noisy individual components of the TDVP equation \cite{Schmitt2020}, the fluctuations have important implications for the time evolution.
One immediate practical consequence is that no higher-order ODE solver is available. Heun's method remains as the typical choice to incorporate adaptive time step sizes imposed by the non-linearity, but it is limited by the Monte Carlo noise.
Moreover, it can be advantageous to dynamically adapt the number of utilized Monte-Carlo samples over time so as to keep the signal-to-noise ratio under control.

Overall, this noisy character of the TDVP also implies a natural maximal time scale up to which the solution can at most remain accurate.
Assuming that the noise is approximately independent from one time step to the next in the sense of white noise and given a noise strength of the order $1/\sqrt{N_s}$ with $N_s$ the number of samples, one can generally expect that the solution is approximated well until a time $t \propto N_s$.
In light of a typical number of $N_s \approx 10^5$ this potential error source has, however, not been of great relevance to date.

\paragraph{Ill-conditioned quantum geometric tensor}
The neural networks used in the NQS approach often exhibit (approximate) redundancies of variational parameters, which is reflected in a severe ill-conditionedness of the QGT. It is common to observe eigenvalue spectra, which are dense across all numerical orders of magnitude \cite{Schmitt2020,Hofmann2022,Donatella2023}. Therefore, integrating TDVP equations such as \eqref{eq:tdvp_1} and \eqref{eq:tdvp_2}, which has the task of solving the corresponding ill-conditioned linear system with stochastic sampling noise at its core, requires careful regularization \cite{Schmitt2020,jVMC2022, Medvidovic2023}. 

Furthermore, the ill-conditionedness can have drastic consequences for the accuracy of the solution of the noisy TDVP equation. 
Due to error propagation, the error of the parameter update $\dot{\vec\theta}$ obtained by solving the equation is proportional to $\kappa/\sqrt{N_s}$, where $\kappa$ is the condition number and $N_s$ is the number of samples \cite{Golub2013}. This issue is commonly addressed by using regularized pseudo-inverses, but accurate solutions may only be obtained with a limited effective reduction of the condition number. Therefore, a seemingly large enough sample size for accurate estimates of the QGT and the force vector, may by far be insufficient to reliably solve the TDVP equation.
While in principle this potentially uncontrolled amplification of Monte-Carlo errors doesn't appear when using the MinSR approach \cite{AoChen2024} to solve the linear system, this is also not practical as MinSR has a computational complexity $\mathcal{O}(N_s^3)$ and typically for dynamics a large number of samples $N_s$ is required.

\paragraph{Towards large NQS}
State-of-the-art approaches to solve TDVP equations rely on the full diagonalization of the QGT for regularization purposes \cite{Schmitt2020,jVMC2022,Medvidovic2023,Nys2024}. This means that the complexity scales cubically with the number of parameters, which severely limits the feasible network sizes. It is an open question, how local optimization approaches can be extended to larger NQS sizes.

\subsubsection{Global optimization}
\label{subsubsec:open_questions_global}
\paragraph{Non-convex gradient-based optimization}
By construction, global approaches to the TDVP pose the known challenges of optimization in non-convex cost landscapes. In particular, the gradient-based optimization can get stuck in local minima or saddle points with insufficient accuracy. Since every simulation time step requires (at least) one minimization of a non-convex cost function, it is crucial to employ suited adaptive schemes that ensure convergence in every time step. Recent works emphasize, that employing (approximations of) natural gradient descent seems necessary to avoid trapping in saddle points \cite{Nys2024, Gravina2024}. Since this requires solving a linear system involving the QGT analogous to the TDVP equations \eqref{eq:tdvp_1} and \eqref{eq:tdvp_2}, the natural gradient approaches come at the cost of dealing with a number of challenges similar to those discussed in Section \ref{subsubsec:open_questions_infintesimal}. As the iterative gradient descent is, however, less susceptible to deviations in individual optimization steps, one may resort to simplifications like a block-diagonal approximation of the geometric tensor \cite{Heskes2000,Nys2024,Zhang2024} or a tangent kernel approach (generalizing MinSR) \cite{Bernacchia2018,Zhang2019,AoChen2024,Zhang2024,Gravina2024}.
Ultimately, it is crucial to achieve very low infidelities, because the deviation of physical observables is only bounded by the square root of the infidelity.
The lack of convergence guarantees in non-convex optimization furthermore implies, that conservation laws or symmetries often cannot be imposed in this approach, which can lead to unfavorable error propagation.
This is well known in general for numerical solutions of ODEs, where typically solvers are desired which automatically respect symmetries.

\paragraph{Monte-Carlo estimation of gradients}
At their core, global optimization approaches rely on the possibility to efficiently compute gradients of the chosen cost function. These have to be formulated as Monte Carlo estimators and it turns out that various choices of the estimator are possible for a given cost function. Since the variances of these estimators can differ strongly, the choice can be crucial for successful optimization \cite{Gravina2024}. Moreover, importance sampling techniques are usually applied to avoid costly sampling from propagated wave functions \cite{Robert2004,Nys2024,Gravina2024}. Overall, it is pivotal to identify the most effective estimation strategies, because all known formulations rely on distributions given by the current NQS during the optimization. This means, that a new sample is required, in principle, in every individual gradient step. Reducing the sampling cost is highly desirable to enhance the efficiency of optimization, which might be for instance achievable using autoregressive networks.

\paragraph{System-size dependent integration time steps}
When targeting a global optimization, it is of key importance to achieve an efficient approximation of the generator $\hat{\mathcal G}$ (cf.~Eq.~\eqref{eq:general_ode}).
When considering Taylor expansions or full generator product expansions to approximate the propagator $\hat\Pi_\tau[\hat{\mathcal G}]=e^{\hat{\mathcal G}\tau}$, it is therefore instructive to consider the series expansion of $\hat\Pi_\tau[\hat{\mathcal G}]$ in terms of the intensive generator density $\hat g\equiv\hat{\mathcal G}/N$ (this is important to isolate the system size dependence of the generator). Any truncation of the series at order $p$ then yields
\begin{align}
    \hat\Pi_\tau[\hat{\mathcal G}]=\sum_{n=0}^p\frac{(N\tau)^n}{n!}\hat g^n+\mathcal O\big((N\tau)^{p+1}\big)\ .
\end{align}
This implies a deviation of $\mathcal O\big((N\tau)^{p+1}\big)$ of the infidelity computed with approximated propagators compared to the actual infidelity, cf.~Eq.~\eqref{eq:infidelity_cost}. In this worst-case scenario, the time step needs to be chosen as $\tau\sim N^{-1}$ for constant accuracy when increasing the system size. 
Clearly, the significance of the errors depends on problem-specific prefactors and time evolution with full generator product expansions has been demonstrated up to seizable system sizes \cite{Gravina2024}. It remains open at which point this property will become prohibitive and whether there are ways of circumventing it. Notice, that the local optimization approach is not affected by this problem, because TDVP equations are derived in the infinitesimal limit $\tau\to0$, where only the lowest order term is relevant by construction.
Overall, it is also an open question to which extent the global optimization for small time steps, as they are necessary for large systems, can yield an advantage as compared to the TDVP equations.

\section{Summary and outlook}

Within recent years the NQS approach has experienced tremendous advances in solving the dynamics of interacting quantum many-body systems.
In this review we have aimed at summarizing the current state of research of the field highlighting both the achievements as well as challenges.
While tensor networks have lifted the theoretical description of quantum matter in one dimension to a new level in the last decades, it still appears as a major challenge in quantum theory to access numerically exact solution of large classes of interacting quantum many-body systems in two and higher dimensions.
This concerns in particular their dynamics where many otherwise powerful methods in quantum physics are facing fundamental limitations, each of which is specific to the utilized method.
For instance, tensor networks suffer from the unavoidable entanglement growth of the matrix contraction complexity, or quantum Monte-Carlo methods are critically limited by their sign problem.
In this light the solution of the dynamics of quantum matter beyond one spatial dimension appears as one of the most promising avenues for the NQS approach.
While some landmark achievements have already been accomplished, further progress in solving quantum real-time evolution with NQS requires to address some of the open question and challenges outlined in Sec.~\ref{sec:challenges}.
Given the still relatively early stages of the field, we anticipate, that further development of the NQS technique will ultimately facilitate important insights for quantum theory by rendering otherwise inaccessible regimes of the dynamical quantum many-body problem solvable.

\paragraph*{Acknowledgements}

This project has received funding from the European Research Council (ERC) under the European Union’s Horizon 2020 research and innovation programme (grant agreement No. 853443), by the German Research Foundation DFG via project 499180199 (FOR 5522) and via project 492547816 (TRR 360), and through the Helmholtz Initiative and Networking Fund, Grant No. VH-NG-1711.

\section{References}
\printbibliography[heading=none]

\end{document}